\documentclass{appolb}
\usepackage{graphicx}
\usepackage{amssymb,eurosym,amsfonts,amsmath}
\usepackage{calc}
\usepackage{array}
\usepackage{multicol}
\usepackage{fancybox}
\usepackage{indentfirst}
\begin{document}
\title{Resonant enlargements of the Poincar\'{e}/AdS (super)algebras from pattern-based analysis
\thanks{Presented at The 8th Conference of the Polish Society on Relativity 2022}%
}
\author{Remigiusz Durka, Krzysztof M. Graczyk
\address{University of Wroc\l{}aw, Institute for Theoretical Physics, pl.\ M.\ Borna 9, 50-204 Wroc\l{}aw, Poland}
\\[3mm]
%{Third Author % of different affiliation
%\address{affiliation}
%}
%\\[3mm]
%the Name(s) of other Author(s)
%\address{affiliation}
}
\maketitle

\begin{abstract}
Applying an efficient pattern-based computational method of generating the so-called 'resonating' algebraic structures results in a broad class of the new Lie (super)algebras. Those structures inherit the AdS base (anti)commutation pattern and can be treated as the enlargements of the Poincar\' {e} or Anti-de-Sitter (super)algebras. Obtained superalgebras are rooted in the semigroup expansion method and Maxwell and Soroka-Soroka algebras, spanned by the Lorentz generator $J_{ab}$, translations $P_{a}$ and additional Lorentz-like generator $Z_{ab}$. Considered configurations include cases up to two fermionic supercharges $Q_{\alpha}$ and offer interesting modiﬁcations to the gauge (super)gravity theories. 
\end{abstract}
  
%%%%%%%%%%%%%%%%%%%%%%%%%%%%%%%%%%%%%%%%%%%%%%%%%%

\section{Introduction}

General Relativity (GR) went through many generalizations and modifications with the exceptionally interesting example of supergravity. In the intermediate step, the GR was formulated as the so-called first-order theory using a set of one-forms: vierbein $e^a$ and spin connection $\omega^{ab}$. In such a framework, in 4D, there is a possible extension to include in the action the Holst term and three topological invariants: Euler, Pontryagin, and Nieh-Yan \cite{Durka:2012wd}. An important example is the MacDowell-Mansouri scheme that
allows one to write GR action in the Yang-Mills style. In 3D one uses the Chern-Simons (CS) action $I_{CS}=\frac{k}{4\pi }\int_{\mathcal{M}}\left\langle \mathbb{A} \wedge d\mathbb{A}+\frac{1}{3} \mathbb{A}\wedge [\mathbb{A},\mathbb{A}]\right\rangle$. It 
offers a convenient framework with straightforward action construction assuring full symmetry invariance. For non-vanishing cosmological constant $\Lambda =\mp \frac{(D-1)(D-2)}{2\ell^{2}}$ and a gauge connection $A=\frac{1}{2}\omega^{ab} J_{ab}+\frac{1}{\ell} e^{a} P_{a}$ 3D gravity reads
\begin{align}
    I^{3D}_{e,\omega}&=\frac{1}{16\pi G}\int \left[ \alpha _{0}\left(
	\omega^{ab}\wedge d\omega _{ab}+\frac{1}{3}\omega^{ab} \wedge \omega
	_{bc}\wedge \omega^{c}{}_{a}+\frac{1}{\ell^2}e_{a} \wedge D_{\omega }e^{a} \right) \right.  \notag \\
	&\left. +\frac{\alpha _{1}}{\ell}\left(\mathcal{R}^{ab}(\omega)\wedge e^{c}+\frac{1}{3\ell^2}e^{a} \wedge e^{b} \wedge e^{c}\right) \epsilon
	_{abc}\right] \,.
\end{align}
We associate with the gauge fields $e^a$ and $\omega^{ab}$,  the corresponding gauge symmetry generators, translations $P_{a}$, and Lorentz generators $J_{ab}$, respectively. Such setup is utilized to directly realize the idea of the supersymmetry between bosons and fermions. Including a fermionic supercharge generator $Q_{\alpha}$ and corresponding spin-$3/2$ Rarita-Schwinger field $\psi^{\alpha}$ leads to the supergravity (SUGRA) theory. Such construction directly results from the underlining structure of the superalgebra that is spanned by the generators of local gauge symmetries: translations $P_a$, Lorentz transformations $J_{ab}$, and supercharge $Q_\alpha$. Note that SUGRA can be constructed starting from the Anti-de Sitter (AdS) setup with negative value $\Lambda = -\frac{3}{\ell^2}$, but there is also the Poincar\'{e}  algebra achieved by setting $\ell\to \infty$ or $\Lambda \to 0$. Supersymmetric extensions of the AdS or the Poincar\'{e} algebras (shown later in the text) are achieved by adding $Q_{\alpha}$ to $J_{ab}$ and $P_{a}$ subalgebra.

Further development of SUGRA was realized by extending the fermionic sector to an arbitrary $\mathcal{N}$ number of fermionic generators $Q^I_{\alpha}$. This paper reviews the exploration of similar ideas. Namely, we focus on the bosonic sector by considering additional bosonic Lorentz-like and translational-like generators. To this end, we shall shortly discuss our recent achievements in the explorations of the superalgebras within the so-called resonant algebra framework. The basic idea originates from the papers of Bacry \cite{Bacry:1970ye} and Schrader~\cite{Schrader:1972zd}. In the later works, it found its natural continuation by D. Soroka and V. Soroka \cite{Soroka:2006aj, Soroka:2011tc}. Eventually, in Refs. \cite{Izaurieta:2006zz, Salgado:2013eut} it was incorporated within generalized contractions that led to the formulation of the Semigroup expansion (S-expansion) procedure. Recently, the resonant approach has been adapted and developed in the series of papers by Durka \textit{et al.} \cite{Durka:2016eun, Durka:2019vnb, Concha:2020atg}. In these works, the primary focus was searching for all possible algebras of a particular type at the given (extended) generator content.

We highlight that the problem of searching for resonant algebras and, more importantly, the superalgebras is quite complicated from the computational point of view. Therefore there was a need to develop the symbolic-computational algorithm \cite{Durka:2021glk, Durka:2022nih}, which allowed us to explore the mentioned algebraic structures. The formulated algorithm is an excellent example of how modern computational tools are changing the standard approaches to constructing algebraical structures. Indeed, we proved that using advanced pattern-based methods allowed us to significantly optimize the exploration of large spaces of potential algebra candidates and push the task in new exciting directions.

The present paper is organized as follows: in Section~\eqref{sec:2}, we give all the necessary theoretical details. Section~\eqref{sec:3} reviews the computational algorithm we have developed. This section highlights essential advancements that made completing the algebraic search in the finite time possible. In Section \eqref{sec:4}, we comment on the potential applications of resonant (super)algebras. This section also discusses one of the obtained superalgebras and the corresponding supergravity theory.

%%%%%%%%%%%%%%%%%%%%%%%%%%%%%%%%%%%%%%%%%%%%%%%%%%

\section{Resonant/S-expanded (super)algebras}
\label{sec:2}

New bosonic Lorentz-like generators historically originate from studying the deformation of the Poincar\'{e} symmetry \cite{Bacry:1970ye, Schrader:1972zd} that included the presence of the constant electromagnetic field. Such an approach led to the modification of the generator's algebra, in which the commutator of the two translations equals the new Lorentz-like generator $Z_{ab}$. This type of algebraic structure is called the Maxwell algebra, and it reads
\begin{align}
\left[ J_{ab},J_{cd}\right] &=\eta_{bc}J_{ad}-\eta_{ac}J_{bd}+\eta_{ad}J_{bc}-\eta_{bd}J_{ac}\,,\notag\\
\left[ J_{ab},P_{b}\right] &=\eta _{bc}P_{a}-\eta _{ac}P_{b}\,,\qquad\left[ P_{a},P_{b}\right] =Z_{ab}\,.
\end{align} 
Recently a similar algebra was introduced in Ref.~\cite{Soroka:2006aj, Soroka:2011tc} but with more complicated structure, namely,
\begin{align}
\left[ J_{ab},J_{cd}\right] &=\eta_{bc}J_{ad}-\eta_{ac}J_{bd}+\eta_{ad}J_{bc}-\eta_{bd}J_{ac}\,,\notag\\
\left[ J_{ab},P_{b}\right] &=\eta _{bc}P_{a}-\eta _{ac}P_{b}\,,\qquad\left[ P_{a},P_{b}\right] =Z_{ab}\,,  \notag\\
\left[Z_{ab},Z_{cd}\right] &=\eta_{bc}Z_{ad}-\eta_{ac}Z_{bd}+\eta_{ad}Z_{bc}-\eta_{bd}Z_{ac}\,,\notag\\
\left[ Z_{ab},P_{b}\right] &=\eta _{bc}P_{a}-\eta _{ac}P_{b}\,.
\end{align}
The above algebras were the subject of various studies, including the supersymmetric versions (for review, see \cite{Durka:2021glk, Durka:2022nih}) and found an interesting continuation in the problem of the generalized In\"{o}n\"{u}-Wigner contractions.

Standardly we connect the (A)dS case (described by a cosmological constant $\Lambda=\mp 3/\ell^2$) with the Poincar\'{e} one ($\Lambda=0$) by introducing parameter $\ell$ that rescales the translation generator:
$P_{a} \to \ell \tilde{P}_{a}\,$. For the AdS case (necessary for the SUGRA formulation) with $\tilde{J},\tilde{P}$ we write explicitly $[\ell \tilde{P}_a, \ell \tilde{P}_b]=\tilde{J}_{ab}$. When this relation is divided by $\ell^2$ and the limit $\ell \to \infty$ is taken, then we reproduce Poincar\'{e} algebra with $[\tilde{P}_a,\tilde{P}_b]=0$. 

If, instead of scalar rescaling, the algebra generators are multiplied by corresponding elements $\lambda_i$ of some semigroup, we have 
\begin{align}
P_{a} \to \lambda_1 \bigotimes \tilde{P}_{a}\qquad J_{ab} \to \lambda_0 \bigotimes \tilde{J}_{ab} \qquad Z_{ab} \to \lambda_2 \bigotimes \tilde{J}_{ab}\,,
\end{align}
and we obtain so-called S-extended algebra denoted as $S \bigotimes \mathfrak{G}$, where $S=\{\lambda_0,\lambda_1,\lambda_2,\ldots\}$ and $\mathfrak{G}=AdS$. However, we can not take the limit discussed above in this scenario. Instead, we have to consider the result of the multiplication between semigroup elements, which ultimately rescales the base AdS generators $\tilde{J}_{ab}, \tilde{P}_{a}$. Finding a new algebra then reduces to finding an existing semigroup. 

Interestingly, formalizing the semigroup expansion delivered a consistent way of finding other algebraic examples with more generators (see paper \cite{Salgado:2013eut}, where the main focus was only on one particular $\mathfrak{C}_{m}$ algebraic family). Also, dealing with "zeros" in the outcomes of the (anti)commutation relations were limited by using a special procedure of $0_S$-reduction. Eventually, trying to achieve the most generality, especially for supergravity, the notion of the resonant algebras was introduced \cite{Durka:2021glk, Durka:2022nih}. Our approach is not based on finding semigroups but rather on a 'smart' brute-force algorithm that directly generates algebras while still adopting some semigroup notation and conventions. It gives a more critical role to the Lorentz generator and deals with zeros universally. The name 'resonant'  comes from the adopting the resonant decomposition of the semigroup elements
\begin{equation}
\lambda_{even}\cdot\lambda_{even}=\lambda_{even}\,, \qquad \lambda_{even}\cdot\lambda_{odd}=\lambda_{odd}\,, \qquad \lambda_{odd}\cdot\lambda_{odd}=\lambda_{even}\,, \label{Eq:lambdas_rules}  
\end{equation}
which reflects the fact of the decomposition of the original AdS algebra
\begin{equation}
[\tilde{J},\tilde{J}] \sim \tilde{J}\,, \qquad [\tilde{J},\tilde{P}] \sim \tilde{P}\,, \qquad [\tilde{P},\tilde{P}] \sim \tilde{J}\,. 
\end{equation}
Let us now discuss supersymmetry. The whole resonant framework can be redefined differently. Namely, adding the new generators must be realized by maintaining the same form of the structure constants as base algebra. Therefore the 'resonant' framework can be understood as simply filling out all the possibilities depending on the (extended) generator's content:
\begin{align}
	\left\{
	\begin{aligned} 
		\left[ \framebox(10,10){~~}_{ab},\ \framebox(10,10){~~}_{cd}\right] & =\eta_{bc} \framebox(10,10){~~}_{ad}-\eta_{ac}\ \framebox(10,10){~~}_{bd}-\eta_{bd} \framebox(10,10){~~}_{ac}+\eta_{ad} \framebox(10,10){~~}_{bc}\, ,\\
		\left[\framebox(10,10){~~}_{ab},\ \dashbox(10,10){~~}_{c}\right] & =\eta_{bc} \dashbox(10,10){~~}_{a}-\eta_{ac} \dashbox(10,10){~~}_{b}\, , \\
		\left[\dashbox(10,10){~~}_{a},\ \dashbox(10,10){~~}_{b}\right] & =\framebox(10,10){~~}_{ab}\, ,\\
		\left[ \framebox(10,10){~~}_{ab},\ovalbox{\phantom{I}}_{\alpha }\right]
		&=\frac{1}{2}\,\left( \Gamma _{ab}\right) _{\alpha }^{\beta }\ovalbox{\phantom{I}}_{\beta}\,,\\ 
		\left[ \dashbox(10,10){~~}_{a},\ \ovalbox{\phantom{I}}_{\alpha }\right]&=\frac{1}{2}\,\left( \Gamma _{a}\right) _{\alpha }^{\beta }\ovalbox{\phantom{I}}_{\beta}\,, \\ 
		\left\{ \ovalbox{\phantom{I}}_{~\alpha },\ \ovalbox{\phantom{I}}_{~\beta }\right\} &=-\left(	\Gamma ^{a} C\right) _{\alpha \beta } \dashbox(10,10){~~}_{~a}+\frac{1}{2}\left(\Gamma ^{ab} C\right) _{\alpha \beta } \framebox(10,10){~~}_{~ab}\,.
	\end{aligned} 
	\right.
	\label{Eq:ALgebraStructure}
\end{align}
Available generators to fill the corresponding entries correspond to
\begin{align}
	& \framebox(10,10){~~}_{~ab} \to J_{ab}, Z_{ab}, \ldots \quad & \dashbox(10,10){~~}_{~a}\to P_{a}, U_{a}, \ldots \quad\quad\, & \ovalbox{\phantom{I}}_{~\alpha } \to Q_\alpha, Y_\alpha, \dots
\end{align}\newpage
\noindent and can be carried out further to include even more extended generator content with more Lorentz-like and translational-like generators as well as supercharges $Q_{\alpha}^I$ (e.g., $Q_{\alpha}^1=Q_{\alpha}$, $Q_{\alpha}^2=Y_{\alpha}$, and so on).

%%%%%%%%%%%%%%%%%%%%%%%%%%%%%%%%%%%%%%%%%%%%%%%%%%

\section{Pattern-based algorithm}
\label{sec:3}

The brute-force approach applied to \eqref{Eq:ALgebraStructure} would correspond to a huge search space. In the case of $JPZU+QY$, it would be unavoidable to check $35$ unique super-Jacobi identities with six (super)commutator substitutions for each of $344\,373\,768$ possible algebra candidates. To deal with this problem, we propose using advanced computer methods within three stages: (a) 'smart' generating algebraic candidates, (b) using pattern search instead of evaluating full Jacobi identities, and (c) applying a stochastic algorithm that 'learns' a type of the Jacobi relation which are the most critical to check. In the first step, we generate the possible candidates, each representing the unique configuration of the generators obeying \eqref{Eq:ALgebraStructure} and written in the form of tables (presented later in the text). We introduce several constraints from the start rejecting obvious incorrect or unphysical situations. For instance, we demand that $(J, X_i) \sim X_i$, for any generator $X_i$. Eventually, when we consider the second fermionic charge $Y_{\alpha}$ we allow for $\{Q_{\alpha}, Y_{\beta}\} =0$ and $\{Y_{\alpha}, Y_{\beta}\}=0$  but require $\{Q_{\alpha}, Q_{\beta}\}\neq 0$.

The algebra is properly defined if the Jacobi relations are satisfied, namely\\[-15pt] \begin{align}
\label{Eq:Jacobi}
0 \equiv &\left( \left(X_i, X_j\right), X_k\right)\oplus \left(\left( X_j, X_k\right), X_i \right) \oplus \left(\left( X_k, X_i\right), X_j\right)\,,
\end{align}
where $\left( .\,,.\right)$ corresponds to either $\{.\,,.\}$ or $[. \, , .]$;  $\oplus$ sign accounts for the possible $(-1)$ factor in the graded Jacobi relation, and $X_{i}, X_{j}, X_{k}$ refers to any generators of the (super)algebra. Note that the enlarged resonant algebra inherits the constant structure content from the base AdS structure \eqref{Eq:ALgebraStructure}. The main idea of our pattern-based approach for searching new possible resonant (super)algebras is based on the observation that  \eqref{Eq:Jacobi}  vanishes because of encoded relations between AdS structure constants. When one considers the resonant framework, where besides new supercharges, one includes new $J$-like, $P$-like generators, the Jacobi identity is satisfied only if the pattern
\begin{equation}
\label{Eq:Pattern}
    Pattern = { \left( \left(X_i, X_j\right), X_k\right),\,\left(\left( X_j, X_k\right), X_i \right) ,\, \left(\left( X_k, X_i\right), X_j\right) }\,,
\end{equation}
leads to the three instances of the same object, i.e., $Pattern = L, L, L$, where $L$ refers to the $0$, $X_i$, or linear combination of $X_i$ (see details in \cite{Durka:2021glk}).

Ultimately, to accelerate the process, we applied an algorithm that 'learns' which types of Jacobi identities are the most critical, i.e., those for which the Jacobi test fails. Further verification starts with checking these critical cases. Note that this process is dynamic, and critical cases are identified during the whole process.

For more complicated scenarios, for instance, the one with additional bosonic $U(1)$ generator  $T$ ($\mathcal{N}=2$ supergravity), the pattern can take a more complicated form \cite{Durka:2022nih}. However, verifying the Jacobi identity can still be reduced to checking the validity of the $Pattern$ structure instead of directly evaluating all expressions.

%%%%%%%%%%%%%%%%%%%%%%%%%%%%%%%%%%%%%%%%%%%%%%%%%%

\section{Results and Conclusions}
\label{sec:4}

Following the program of finding all (super)algebra possibilities for a given generator content, we arrive at the following number of examples \cite{Durka:2021glk}:
\\[-5px]
\begin{table}[htb]
	\label{table-superalgebras}
	\begin{center}
		\begin{tabular}{l|l|l}
			~1$\times$ J             & ~1$\times$ J+Q             & ~~0$\times$ J+QY             \\ \hline
			~2$\times$ JP            & ~2$\times$ JP+Q            & ~10$\times$ JP+QY    \\ \hline
			~6$\times$ JPZ    & ~9$\times$ JPZ+Q  & 102$\times$ JPZ+QY  \\ \hline
			30$\times$ JPZU & 43$\times$ JPZU+Q &667$\times$ JPZU+QY
		\end{tabular}
		%\caption[]{The number of resonant (super)algebras depending on used generators.}
	\end{center}
\end{table}\\[-15px]
Within the plethora of results shown above, some of them represent the 'exotic' cases, where $\{Q, Q\}\neq 0$ does not give translations $P_a$. The $\mathcal{N}=1$ resonant $JPZ+Q$ cases have already been discussed in~\cite{Concha:2020atg}. 

All the obtained algebra outputs in our works are presented in commutation tables, which provide a concise format that immediately highlights all the differences between algebras. Such a form also emphasizes the independence of the structure constants, so depending on a particular application, we can use either notation of AdS structure constants from $4D$ case or $3D$. For other situations, one must consider using properly defined base algebra and commutation relation templates of the structure constants and a necessary number of generators. Below we schematically write the tables for the super Poincar\'{e} and AdS to show their structure while keeping in mind all the coefficients, gammas, Etc.
\begin{align}\label{Poincare}
\begin{tabular}[t]{c|cc}
		\lbrack .,.] & $J$ & $P$ \\ \hline
		$J$ & $J$ & $P$ \\ 
		$P$ & $P$ & $0$
	\end{tabular}
	\qquad 
	\begin{tabular}[t]{c|c}
		\lbrack .,.] & $Q$ \\ \hline
		$J$ & $Q$ \\ 
		$P$ & $0$
	\end{tabular}
	\qquad 	
	\begin{tabular}[t]{c|c}
		\{.,.\} & $Q$ \\ \hline
		$Q$ & $P$
	\end{tabular}
\end{align}\\[-20px]
\begin{align}
	\begin{tabular}[t]{c|cc}\label{AdS}
		\lbrack .,.] & $J$ & $P$ \\ \hline
		$J$ & $J$ & $P$ \\ 
		$P$ & $P$ & $J$
	\end{tabular}
	\qquad 
	\begin{tabular}[t]{c|c}
		\lbrack .,.] & $Q$ \\ \hline
		$J$ & $Q$ \\ 
		$P$ & $Q$
	\end{tabular}
	\qquad 
	\begin{tabular}[t]{c|c}
		\{.,.\} & $Q$ \\ \hline
		$Q$ & $P+J$
	\end{tabular}
\end{align}

Certainly, we are more interested in building gravity models than the algebras themselves. Notable examples of gravity models correspond to theories: $\mathcal{N}=0$ (GR), $\mathcal{N}=1$ (SUGRA) and $\mathcal{N}=2$ with two supercharges $Q_\alpha$ and $Y_\alpha$. The second fermionic supercharge offers a possibility of bi-supergravity in analogy to bi-metric theory. With the two copies of Lorentz-like and translational-like generators, we get $30$ bosonic algebra cases. If $J_{ab},P_a \to \omega^{ab}, e^a \to g_{\mu\nu}$ we might consider $Z_{ab},U_a \to h^{ab}, k^a \to h_{\mu\nu}$. Instead of arbitrary mixed terms, we have $30$ explicit theories with distinct Lagrangians due to the corresponding $30$ algebras. 

As an example, let us show how starting from much simpler superalgebra than discussed in \cite{Durka:2021glk} (denoted as $\mathcal{B}_5$ with single factor $\ell$ in a definition of the connection) allows us to construct the corresponding bi-supergravity Lagrangian. After using 3D dual definitions of fields ($\omega^a$ and $h^a$) and generators ($J_a$ and $Z_a$) we rewrite the $\mathcal{B}_5$ connection as $\mathbb{A}=\omega^{a}J_{a}+\frac{1}{\ell} e^{a}P_{a}+ h^{a}Z_{a}+\frac{1}{\ell} k^{a} U_{a}+\frac{1}{\sqrt{\ell}}\psi^{\alpha}Q_{\alpha}+\frac{1}{\sqrt{\ell}}\chi^{\alpha}Y_{\alpha}$ to obtain $\mathbb{F}=F^{a}J_{a}+\frac{1}{\ell} T^{a}P_{a}+ H^{a}Z_{a}+\frac{1}{\ell} K^{a} U_{a}+\frac{1}{\sqrt{\ell}}\mathcal{F}^{\alpha}Q_{\alpha}+\frac{1}{\sqrt{\ell}}\mathcal{G}^{\alpha}Y_{\alpha} $ with the following components
\small
\begin{align}
	F^{a}& =\mathcal{R}^{a}(\omega)+\frac{1}{2\ell^2}\epsilon^{abc} e_b e_c+\frac{1}{2\ell}\,\bar{\psi}\Gamma^{a}\psi\,,  \notag \\
	T^{a}& =D_\omega e^{a}+\frac{1}{2}\bar{\psi}\Gamma^{a}\psi \,,  \qquad 	\mathcal{F}=\mathcal{D}_{\omega }\psi +\frac{1}{2\ell}\,e^{a}\Gamma_{a}\psi  \,,  \notag\\
	H^{a}& =D_{\omega}h^{a}+\frac{1}{\ell^2}\epsilon^{abc}e_b k_c+\frac{1}{2\ell}\,\bar{\psi}\Gamma^{ab}\chi+\frac{1}{2\ell}\,\bar{\chi}\Gamma^{ab}\psi\,,  \notag \\
	K^{a}& =  D_{\omega} k^{a}+\epsilon^{abc}h_b e_c+\frac{1}{2}\,\bar{\psi}\Gamma^{a}\chi+\frac{1}{2}\,\bar{\chi}\Gamma^{a}\psi\,,  \notag \\
	\mathcal{G} & =\mathcal{D}_{\omega }\chi +\frac{1}{2\ell}e^{a}\Gamma _{a}\chi +\frac{1}{2} \,h^{a}\Gamma _{a}\psi+\frac{1}{2\ell}\,k^{a}\Gamma_{a}\psi \,.
\end{align}\normalsize
The corresponding Lagrangian for 3D Chern-Simons model reads:
\small
\begin{eqnarray}
\mathcal{L}^{\mathcal{B}_{5}}&=&\alpha _{0}\left(
	\omega ^{a}d\omega _{a}+\frac{1}{3}\epsilon^{abc}\omega _{a}\omega
	_{b}\omega _{c}+\frac{1}{\ell^2}e_{a}D_{\omega }e^{a}-\frac{2}{\ell}\left(
	\bar{\psi}\mathcal{F}+
	\bar{\chi}\mathcal{G}\right)\right)  \notag \\
	&&+\alpha _{1}\left( \frac{2}{\ell}e_{a}\mathcal{R}^{a}+\frac{1}{3\ell^3}\epsilon^{abc}e_{a}e_{b}e_{c}+\frac{2}{\ell} \left(\bar{\psi}\mathcal{F}+
	\bar{\chi}\mathcal{G}\right) \right)  \notag  \\
	&& +\alpha _{2}\left(2h_{a}\mathcal{R}
	^{a}+\frac{2}{\ell^2}e_a D_\omega k^a+\frac{1}{\ell^2}\epsilon^{abc}e_{a}e_{b}h_{c}-\frac{2}{\ell} \left(\bar{\psi}\mathcal{G}+\bar{\chi}\mathcal{F}\right) \right)  \notag  \\
	&&+\alpha _{3}\left(\frac{2}{\ell}k_{a}\mathcal{R}
	^{a}+\frac{2}{\ell}e_{a}D_{\omega }h^{a}+\frac{1}{\ell^3}\epsilon^{abc}e_{a}e_{b}k_{c}+\frac{2}{\ell}\left(\bar{\psi}\mathcal{G}+\bar{\chi}\mathcal{F} \right)\right)  \,.
\end{eqnarray}\normalsize
Notice the appearance of the sub-invariant sectors introduced by the arbitrarily valued constants $\alpha_i$.

Interestingly, $\mathcal{N}=2$ case allows for including a notion of the central charges or internal symmetry generator $T$ giving rise to having $U(1)$ $A_{\mu}$ field incorporated in the algebraic framework. The procedure caused some additional complications to the algorithm and the use of a modification of the expected super Jacobi template. For $JPZ+QY+T$ configuration \cite{Durka:2022nih}, we found only a single algebraic realization through our analysis.

Note that typically $[P_a, Q^I] \sim Q^I$ or $0$, but in Ref. \cite{Durka:2021glk} we report ten $JP+QY$ configurations (with two supercharges but without $T$ generator), with $[P_a, Q^I]~Q^{J}$. Algebras with such features, according to our knowledge, are not studied in the literature as with two supercharges, one immediately transits to the configurations with the central charge or internal symmetry.

The same method can be applied to other base structures. The starting superalgebras are different in other dimensions, but keeping the same structure constants while adding generators will lead to similar generalization.

The resonant algebra framework and corresponding computational methods proved to be helpful in better understanding of constructing the (super)gravity theories. In the future, one might consider a more non-trivial closing of the resonant algebras, for instance, focusing on breaking $\{Q, Q\}=\{Y, Y\}$ condition, which we already explored in the first paper \cite{Durka:2021glk}.\\

\noindent \textbf{Acknowledgments}: \textit{The project for both authors was partly supported by the ''Excellence initiative - research university'' for the years 2020-2026 for the University of  Wroc\l{}aw.}\\[-15pt]

%%%%%%%%%%%%%%%%%%%%%%%%%%%%%%% Bibliography 

\end{document}